\documentclass[11pt,dvips]{article}
\usepackage{graphicx}
\usepackage{amsmath}
\usepackage{amssymb}
\usepackage{latexsym}
\usepackage{color}

\begin{document}

\thispagestyle{empty}

\begin{center}
\vspace{1.8cm}

{\Large \textbf{Entanglement versus Gaussian quantum discord in a
double-cavity opto-mechanical system}}


\vspace{1.5cm}

\textbf{J. El Qars}$^{a}${\footnote{%
email: \textsf{j.elqars@gmail.com}}}, \textbf{M. Daoud}$^{b,c,d}${\footnote{%
email: \textsf{m$_-$daoud@hotmail.com}}} and \textbf{Ahl Laamara}$^{a,e}$ {%
\footnote{%
email: \textsf{ahllaamara@gmail.com}}}

\vspace{0.5cm}

$^{a}$\textit{LPHE-MS, Faculty of Sciences, University Mohammed V, Rabat,
Morocco}\\[0.5em]
$^{b}$\textit{Max Planck Institute for the Physics of Complex Systems,
Dresden, Germany}\\[0.5em]
$^{c}$\textit{Abdus Salam International Centre for Theoretical Physics,
Miramare, Trieste, Italy}\\[0.5em]
$^{d}$\textit{Department of Physics , Faculty of Sciences, University Ibnou
Zohr, Agadir, Morocco}\\[0.5em]
$^{e}$\textit{Centre of Physics and Mathematics (CPM), University Mohammed
V, Rabat, Morocco}\\[0.5em]

\vspace{3cm} \textbf{Abstract}
\end{center}

\baselineskip=18pt \medskip

In this paper we investigate the robustness of the quantum correlations
against the environment effects in various opto-mechanical bipartite
systems. For two spatially separated opto-mechanical cavities, we give
analytical formula for the global covariance matrix involving two mechanical
modes and two optical modes. The logarithmic negativity as an indicator of
the degree of entanglement and the Gaussian quantum discord which is a
witness of quantumness of correlations are used as quantifiers to evaluate
the different pairwise quantum correlations in the whole system. The
evolution of the quantum correlations existing in this opto-mechanical
system are analyzed in terms of the thermal bath temperature, squeezing
parameter and the opto-mechanical cooperativity. We find that with desirable
choice of these parameters, it is possible either enhance or annihilate the
quantum correlations in the system. Various scenarios are discussed in
detail.

\newpage

\section{Introduction}

Quantum correlations transfer between light and matter is currently viewed
as a key ingredient for future applications in the context of quantum
communications and information processing \cite{Tian,AClerk23,Singh,Palomaki}%
. Storing information in the matter degree of freedom is preferable to
overcome the difficulties of storage and localization encountered with
photons. In this context, over the last two decades, the transfer of quantum
correlations from photons to matter has raised widespread interest from a
purely theoretical point of view supported by significant experimental
achievements. In fact, the opto-mechanical coupling between the
electromagnetic mode in a quantum cavity and the mechanical motion of a
nano-mechanical resonator by exploiting the radiation pressure force offers
a platform to explore the entanglement transfer between light and matter.
The opto-mechanical systems provide also very promising tools to create and
manipulate entanglement at mesoscopic scales. The appropriate setup
extensively used in investigating quantum correlations in opto-mechanical
systems, and subsequently to understand the entanglement transfer between
optical and mechanical modes, is the Fabry-Perot cavity \cite%
{Gigan,Bouwmeester}. Indeed, various schemes using Fabry-Perot cavity were
reported in the literature from several perspectives and for different
purposes \cite{Vitali 2002,
Peng,PinardVitali4,VitaliVedral17,RhobadiSimon,PaternostroChiara,A.Mari,Hartmann,PaternostroVitaliAspelmeyer,Tiantian,Sete et al}%
. Clearly, the increasing interest in transferring the quantum correlations
from microscopic systems to mesoscopic ones is primarily motivated by the
use of non-classical entangled states of continuous variable systems quantum
information processing, communication and computation. Different measures to
quantify the degree of intricacy in bipartite quantum systems were discussed
in the literature. In particular, for a long, time the entanglement \cite%
{wootters} has been regarded as the key ingredient to distinguish between
entangled and separable states and subsequently between the quantum and
classical correlations. In this picture, separability has been often
identified with the absence of quantum correlations. However, now it is well
established that quantum correlation can be present in separable states.
Indeed, the notion of quantum discord, introduced in \cite%
{OllivierZurek6,Vedral7}, which goes beyond the entanglement, is the
appropriate measure to deal with the quantum correlations in bipartite
quantum systems, especially the ones those prepared in mixed states. The
quantum discord, originally defined and evaluated for finite dimensional
system, was extended to the domain of continuous variable systems and
especially in analyzing the bipartite quantum correlations in Gaussian mode
states\ \cite{GeordaParis9,AdessoDatta10,Olivares}. \newline
In this paper, to quantify the degree of quantum correlations, we shall use
the logarithmic negativity and the Gaussian quantum discord. We stress that
the characterization of quantum correlations in opto-mechanical systems is
essential to understand the transfer of (quantum correlations from light to
matter) entanglement between optical and macroscopic vibrational modes. We
notice that other measures and criteria were used in this sense. One may
quote for instance, Duan and Simon entanglement criterion proposed
simultaneously and independently by Duan et al \cite{Duan} and Simon \cite%
{Simon} which provide the inseparability condition of two continuous
variable systems, the Mancini separability criterion \cite{Vitali 2002}
which is valid for any state of any bipartite system and generalizes the
already mentioned criteria. The logarithmic negativity \cite%
{VidalWerner30,Adesso et al32} was also used to quantify the amount between
two Gaussian modes. However, this measure is not sufficient to specify
completely the quantum correlations present in the system, especially for
mixed states. Henceforth, the appropriate measure in this case is the
Gaussian quantum discord \cite{AdessoDatta10} ( see also \cite{GeordaParis9}%
). In fact, this measure has been shown useful in determining the
non-classical correlations between two spatially distant mechanical
oscillators \cite{Mauro21}. In this work, we investigate the non-classical
correlations between the different modes in an opto-mechanical system
consisting of two movables mirrors of two spatially separated Fabry-Perot
cavities. Each cavity is pumped by a squeezed light. A complete description
of this opto-mechanical model is provided in section II. We give the
corresponding Hamiltonian. We solve the associated quantum Langevin
equations to determine the explicit form of the global covariance matrix
involving the quadratures of mechanical and optical modes. In section III,
using the logarithmic negativity, we investigate the separability between
the different modes in model. A special emphasis is devoted in section IV to
the situation where the logarithmic negativity is zero. In this case, the
Gaussian quantum discord is used to quantify the quantum correlations appear
beyond entanglement. Concluding remarks close this paper.

\section{ System and Hamiltonian}

\subsection{ The system}

\begin{figure}[tbh]
\centerline{\includegraphics[width=7cm]{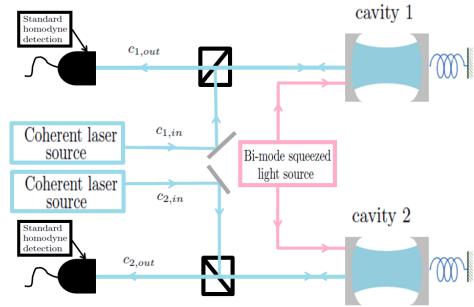}}
\caption{Schematic of two identical opto-mechanical Fabry-Perot cavities
which are pumped by identical laser fields (power $\mathrm{P}$ and frequency
$\protect\omega _{L}$) and two-mode squeezed light, generated for example by
the spontaneous parametric down conversion source (SPDC). Each movable
mirror is treated as a mechanical oscillator characterized respectively by
the frequency $\protect\omega _{\protect\mu }$ and the damping rate $\protect%
\gamma $. }
\label{doublecavite}
\end{figure}

The opto-mechanical system considered in this paper, consists of two
identical Fabry-Perot cavities (see Fig.1). Each cavity is composed by two
mirrors. The first mirror is fixed and partially transmitting, the second is
movable and perfectly reflecting. As depicted in Fig.1, each cavity is
pumped simultaneously by coherent laser field and squeezed light produced by
using either the SPDC source (spontaneous parametric down-conversion) \cite%
{Burnham11,Shih12} or by techniques of nonlinear optic \cite%
{Yurke13,Shahriar14}. $\varepsilon =\sqrt{\frac{2\kappa P}{\hbar \omega _{L}}%
}$ is the amplitude of the pump laser, where the parameter $\kappa $ denotes
the energy decay rate of the two cavities, $\omega _{L}$ and $P$ are
respectively the frequency and the power of the external laser sources. The
opto-mechanical coupling via the radiation pressure \cite{Karrai15} between
the cavity field and the movable mirror is characterized by the coefficient $%
g$ given by $g=\frac{\omega _{c}}{L}\sqrt{\frac{\hbar }{\mu \omega _{\mu }}}$
with $\omega _{c}$ and $L$ denoting respectively the frequency and the
length of each cavity. Finally, each movable mirror will be treated as a
quantum mechanical harmonic oscillator with the damping rate $\gamma $, the
mass $\mu $ and the frequency $\omega _{\mu }$.

\subsection{ The Hamiltonian}

\noindent In the in a frame rotating with $\omega _{L}$, the Hamiltonian of
the system is given by \cite{CKLaw16}
\begin{equation}
H=\sum_{i=1}^{2}\left( \left( \omega _{c}-\omega _{L}\right) c_{i}^{\dag
}c_{i}+\omega _{\mu }b_{i}^{\dag }b_{i}+gc_{i}^{\dag }c_{i}(b_{i}^{\dag
}+b_{i})+\varepsilon (e^{i\varphi _{i}}c_{i}^{\dag }+e^{-i\varphi
_{i}}c_{i})\right) ,  \label{Hamiltonian}
\end{equation}%
where $c_{i}^{\dag }$ and $c_{i}$ are respectively the creation and the
annihilation operators for the $i^{th}$ optical mode. They satisfy the usual
bosonic commutation relations. Similarly, $b_{i}^{\dag }$ and $b_{i}$ stand
for the creation and the annihilation operators for the $i^{th}$ mechanical
mode. In Eq. (\ref{Hamiltonian}), $\varphi _{i}$ denotes the $i^{th}$ input
laser field phase. To simplify, we assume $\varphi _{1}=\varphi _{2}=\varphi
$. In the Heisenberg representation, the quantum Langevin equations for
optical and mechanical modes read
\begin{eqnarray}
\frac{dc_{i}}{dt} &=&i\left[ H,c_{i}\right] -\frac{\kappa }{2}c_{i}+\sqrt{%
\kappa }c_{i}^{in}=-\left( \frac{\kappa }{2}-i\Delta \right)
c_{i}-igc_{i}(b_{i}^{\dag }+b_{i})-i\varepsilon e^{i\varphi }+\sqrt{\kappa }%
c_{i}^{in},  \label{cdote} \\
\frac{db_{i}}{dt} &=&i\left[ H,b_{i}\right] -\frac{\gamma }{2}b_{i}+\sqrt{%
\gamma }\xi _{i}=-\left( \frac{\gamma }{2}+i\omega _{\mu }\right)
b_{i}-igc_{i}^{\dag }c_{i}+\sqrt{\gamma }\xi _{i},  \label{bdote}
\end{eqnarray}%
where $\Delta =\omega _{L}-\omega _{c}$ is the laser detuning, $c_{i}^{in}$
denotes the $i^{th}$ input squeezed vacuum noise operator, $\xi _{i}$ is the
$i^{th}$ noise operator associated to the Brownian motion of the $i^{th}$
movable mirror. The input squeezed vacuum noise operators $c_{i}^{in}$ have
the following nonzero frequency-domain correlation functions \cite%
{HuangAgarwal19,Gardiner20}
\begin{eqnarray}
\langle c_{i}^{in^{\dag }}(-\omega )c_{i}^{in}(\omega ^{\prime })\rangle
&=&2\pi N\delta (\omega +\omega ^{\prime }),  \label{correZ1} \\
\langle c_{i}^{in}(\omega )c_{i}^{in^{\dag }}(-\omega ^{\prime })\rangle
&=&2\pi (N+1)\delta (\omega +\omega ^{\prime }),  \label{correZ2} \\
\langle c_{1}^{in}(\omega )c_{2}^{in}(\omega ^{\prime })\rangle &=&2\pi
M\delta (\omega +\omega ^{\prime }-2\omega _{\mu }),  \label{correZ3} \\
\langle c_{1}^{in^{\dag }}(-\omega )c_{2}^{in^{\dag }}(-\omega ^{\prime
})\rangle &=&2\pi M\delta (\omega +\omega ^{\prime }+2\omega _{\mu }),
\label{correZ4}
\end{eqnarray}%
with $N=\sinh ^{2}r$ and $M=\sinh r\cosh r$, where $r$ is the squeezing
parameter characterizing the squeezed light. The noise operators $\xi _{i}$
in Eq. (\ref{bdote}) have zero mean value. In general, the mechanical baths
are not Markovian \cite{VitaliVedral17,GiovannettiVitali18}. The mechanical
baths can be considered as Markovian when the mechanical oscillator
frequency $\omega _{\mu }$ is larger than the damping rate $\gamma $. In
this situation, we have the following Markovian delta-correlated relations
\begin{eqnarray}
\langle \xi _{i}^{\dag }(-\omega )\xi _{i}(\omega ^{\prime })\rangle &=&2\pi
n_{\mathrm{th}}\delta (\omega +\omega ^{\prime }),  \label{correBro1} \\
\langle \xi _{i}(\omega )\xi _{i}^{\dag }(-\omega ^{\prime })\rangle &=&2\pi
(n_{\mathrm{th}}+1)\delta (\omega +\omega ^{\prime }),  \label{correBro2}
\end{eqnarray}%
where $n_{\mathrm{th}}=\bigg(\exp \big[\frac{\hbar \omega _{\mu }}{K_{B}T}%
\big]-1\bigg)^{-1}$ is the mean thermal photons number and $T$ is the
mechanical bath temperature. The quadratic terms in Eqs. (\ref{cdote}) and (%
\ref{bdote}) are due essentially to the non-linear nature of the radiation
pressure \cite{Mauro21}. To solve the system of Eqs. (\ref{cdote}) and (\ref%
{bdote}), we define the operators \cite{FabrePinard22}
\begin{equation}
\delta b_{i}=b_{i}-b_{si},\qquad \delta c_{i}=c_{i}-c_{si},
\label{operat-fluct}
\end{equation}%
where $b_{si}$ and $c_{si}$ are the steady-state averages for mechanical and
optical operators respectively. From Eqs. (\ref{cdote}) and (\ref{bdote}),
one can check that they are given by
\begin{equation}
\langle c_{i}\rangle =c_{si}=\frac{-i\varepsilon e^{i\varphi }}{\frac{\kappa
}{2}-i\Delta _{\mathrm{eff}}}\qquad \langle b_{i}\rangle =b_{si}=\frac{%
-ig\left\vert c_{si}\right\vert ^{2}}{\frac{\gamma }{2}+i\omega _{\mu }},
\label{c and b average}
\end{equation}%
with $\Delta _{\mathrm{eff}}=\Delta -g(b_{si}+{\bar{b}_{si}})$ denotes the
effective cavity detuning including the mirrors displacements due to
radiation pressure. Reporting Eq. (\ref{operat-fluct}) in Eqs. ( \ref{cdote}%
) and (\ref{bdote}), the fluctuations $\delta b_{i}$ and $\delta c_{i}$ of
the operators $c_{i}$ and $b_{i}$, around the steady states, obey to the
following equations
\begin{eqnarray}
\delta \dot{c}_{i} &=&-\left( \frac{\kappa }{2}-i\Delta _{\mathrm{eff}%
}\right) \delta c_{i}-G\left( \delta b_{i}^{\dag }+\delta b_{i}\right) +%
\sqrt{\kappa }~c_{i}^{in},  \label{deltacdote} \\
\delta \dot{b}_{i} &=&-\left( \frac{\gamma }{2}+i\omega _{\mu }\right)
\delta b_{i}+G\left( \delta c_{i}-\delta c_{i}^{\dag }\right) +\sqrt{\gamma }%
~\xi _{i},  \label{deltabdote}
\end{eqnarray}%
with $G=g\left\vert c_{si}\right\vert $ is the many-photon opto-mechanical
coupling. In deriving the last evolution equations, we have deliberately
chosen the input field phase to be $\tan \varphi =\frac{-2\Delta _{\mathrm{%
eff}}}{\kappa }$. This is legitimate since the coherent field can be
produced with an arbitrary phase. For this special value of the phase, we
have $c_{si}=-i\left\vert c_{si}\right\vert $. Furthermore, setting $\Delta
_{\mathrm{eff}}=-\omega _{\mu }$, which corresponds to the quantum state
transfer \cite{AClerk23}. Using the rotating wave approximation at frequency
$\omega _{\mu }$.i.e., for each operator $O$, we have $\tilde{O}=O\exp
(i\omega _{\mu }t)$ and we neglect the fast rotating terms,one gets
\begin{equation}
\delta \dot{\tilde{c}}_{i}=-\frac{\kappa }{2}~\delta \tilde{c}_{i}-G~\delta
\tilde{b}_{i}+\sqrt{\kappa }~\tilde{c}_{i}^{in}\quad ,\quad \delta \dot{%
\tilde{b}}_{i}=-\frac{\gamma }{2}~\delta \tilde{b}_{i}+G~\delta \tilde{c}%
_{i}+\sqrt{\gamma }~\tilde{\xi}_{i},  \label{(c,b)dote in RWA}
\end{equation}%
Finally, using the Fourier transform of the last differential equations, the
explicit expressions for $\delta \tilde{c}_{i}$ and $\delta \tilde{b}_{i}$
write
\begin{eqnarray}
\delta \tilde{c}_{i}(\omega ) &=&\frac{-G}{d(\omega )}\sqrt{\gamma }~\tilde{%
\xi}_{i}(\omega )+\frac{\left( \frac{\gamma }{2}+i\omega \right) }{d(\omega )%
}\sqrt{\kappa }~\tilde{c}_{i}^{in}(\omega ),  \label{deltactilde} \\
\delta \tilde{b}_{i}(\omega ) &=&\frac{\left( \frac{\kappa }{2}+i\omega
\right) }{d(\omega )}\sqrt{\gamma }~\tilde{\xi}_{i}(\omega )+\frac{G}{%
d(\omega )}\sqrt{\kappa }~\tilde{c}_{i}^{in}(\omega ),  \label{deltabtilde}
\end{eqnarray}%
with $d(\omega )=G^{2}+\left( \frac{\gamma }{2}+i\omega \right) \left( \frac{
\kappa }{2}+i\omega \right)$.

\subsection{Covariance matrix}

To estimate entanglement and Gaussian quantum discord between different
bipartite modes selected from the global system, we will derive the explicit
formula of the covariance matrix describing the whole system. For this, we
introduce the following quadrature operators (EPR-type quadrature operators
for mechanical and optical modes)
\begin{equation}
\delta X^{\mathrm{m}_{i}}(\omega )=\frac{\delta \tilde{b}_{i}^{\dagger }
\text{ }+\text{ }\delta \tilde{b}_{i}}{\sqrt{2}},\qquad \delta Y^{\mathrm{m}%
_{i}}(\omega )=i\frac{\delta \tilde{b}_{i}^{\dagger }\text{ }-\text{ }\delta
\tilde{b}_{i}}{\sqrt{2}},  \label{QOM}
\end{equation}%
\begin{equation}
\delta X^{\mathrm{o}_{i}}(\omega )=\frac{\delta \tilde{c}_{i}^{\dagger }%
\text{ }+\text{ }\delta \tilde{c}_{i}}{\sqrt{2}},\qquad \delta Y^{\mathrm{o}%
_{i}}(\omega )=i\frac{\delta \tilde{c}_{i}^{\dagger }\text{ }-\text{ }\delta
\tilde{c}_{i}}{\sqrt{2}},  \label{QOO}
\end{equation}%
where $\delta X^{s_{i}}$ and $\delta Y^{s_{i}}$ are respectively the $i^{th}$
($i=1,2$) position and momentum quadrature operators associated to the
mechanical modes Eq. (\ref{QOM})(with $s\equiv \mathrm{m}$) and the optical
modes Eq. (\ref{QOO}) ($s\equiv \mathrm{o}$ ). For continuous variables, it
is appropriate to specify the system within the covariance matrix formalism
\cite{AdessoDatta10,AdessoF24,Braunstein26}. We introduce the 8-component
vector
\begin{equation*}
U^{\mathrm{T}}=(\delta X^{\mathrm{m}_{1}}(\omega ),\delta X^{\mathrm{m}%
_{2}}(\omega ),\delta Y^{\mathrm{m}_{1}}(\omega ),\delta Y^{\mathrm{m}%
_{2}}(\omega ),\delta X^{\mathrm{o}_{1}}(\omega ),\delta X^{\mathrm{o}%
_{2}}(\omega ),\delta Y^{\mathrm{o}_{1}}(\omega ),\delta Y^{\mathrm{o}%
_{2}}(\omega )),
\end{equation*}%
where the subscript $\mathrm{T}$ stands for the transposition operation. The
corresponding covariance matrix elements can be evaluated explicitly by
using the correlations properties of the noise operators $c_{i}^{in}$ and $%
\xi _{i}$ ( Eqs. (\ref{correZ1})-(\ref{correBro2})) and the following
relation \cite{Mauro21}
\begin{equation}
\sigma _{pq}=\frac{1}{4\pi ^{2}}\int_{-\infty }^{+\infty }\int_{-\infty
}^{+\infty }d\omega d\omega ^{\prime }e^{-i(\omega +\omega ^{\prime
})t}\sigma _{pq}(\omega ,\omega ^{\prime }),  \label{CM elements}
\end{equation}%
where the frequency-domain correlation function between the elements $p$ and
$q$ of the vector $U^{\mathrm{T}}$ are defined by
\begin{equation}
\sigma _{pq}(\omega ,\omega ^{\prime })=\frac{1}{2}\langle \left\{
U_{p}(\omega ),U_{q}(\omega ^{\prime })\right\} \rangle ,
\label{Fourier of CM}
\end{equation}%
for $p,q=1,..,8$. \ After some algebra, we finally obtain

\begin{equation}
\sigma =\left(
\begin{array}{cccccccc}
a_{1} & 0 & c_{1} & 0 & c_{3} & 0 & c_{4} & 0 \\
0 & a_{1} & 0 & -c_{1} & 0 & c_{3} & 0 & -c_{4} \\
c_{1} & 0 & a_{1} & 0 & c_{4} & 0 & c_{3} & 0 \\
0 & -c_{1} & 0 & a_{1} & 0 & -c_{4} & 0 & c_{3} \\
c_{3} & 0 & c_{4} & 0 & a_{2} & 0 & c_{2} & 0 \\
0 & c_{3} & 0 & -c_{4} & 0 & a_{2} & 0 & -c_{2} \\
c_{4} & 0 & c_{3} & 0 & c_{2} & 0 & a_{2} & 0 \\
0 & -c_{4} & 0 & c_{3} & 0 & -c_{2} & 0 & a_{2}%
\end{array}%
\right) ,  \label{Global CM}
\end{equation}

\noindent where

\begin{eqnarray}
a_{1} &=&\frac{\beta \cosh 2r}{2(1+\alpha )\left( 1+\beta \right) }+\frac{
\left( 2n_{\mathrm{th}}+1\right) \left( 1+\alpha +\alpha \beta \right) }{
2(1+\alpha )\left( 1+\beta \right) }\text{ \ \ \ \ \ \ \ \ }c_{1}=\frac{
\beta \sinh 2r}{2(1+\alpha )\left( 1+\beta \right) },  \label{system M} \\
a_{2} &=&\frac{\cosh 2r\left( 1+\alpha +\beta \right) }{2(1+\alpha )\left(
1+\beta \right) }+\frac{\left( 2n_{\mathrm{th}}+1\right) \alpha \beta }{
2(1+\alpha )(1+\beta )}\text{ \ \ \ \ \ \ \ \ \ \ \ \ \ \ } c_{2}=\frac{%
\sinh 2r\left( 1+\alpha +\beta \right) }{2(1+\alpha )\left( 1+\beta \right) }%
,  \label{system O} \\
c_{3} &=&\frac{\sqrt{\alpha \beta }}{2\left( 1+\alpha \right) (1+\beta )} %
\Big(-\left( 2n_{\mathrm{th}}+1\right) +\cosh 2r\Big)\text{\ \ \ \ \ \ \ \ \
\ }c_{4}= \frac{\sqrt{\alpha \beta }\sinh 2r}{2\left( 1+\alpha \right)
(1+\beta )},  \label{system H}
\end{eqnarray}

\noindent where $\alpha =\frac{\gamma }{\kappa }$ is the damping ratio \cite%
{Clerk} and $\beta =\frac{4G^{2}}{\kappa \gamma }$ represents the
opto-mechanical cooperativity. This parameter measures the coupling degree
between mechanical and optical modes \cite{Sete et al, PetersonKampel29}.

\section{Entanglement analysis via the logarithmic negativity}

From the covariance matrix given by Eq. (\ref{Global CM}), we shall now
investigate the bipartite entanglement between different modes in the
system. Indeed, we quantify the quantum correlations using the logarithmic
negativity between: the mechanical mode 1 and the mechanical mode 2
(subsystem $\mathrm{(i)}$), the optical mode 1 and the optical mode 2
(subsystem $\mathrm{(ii)}$), the mechanical mode 1 (resp. 2) and the optical
mode 1 (resp. 2) (subsystem $\mathrm{(iii)}$) and finally the mechanical
mode 1 (resp. 2) and the optical mode 2 (resp. 1) (subsystem $\mathrm{(iv)}$%
). For each pair of modes, the corresponding covariance matrix can be
derived from the global covariance matrix (\ref{Global CM}). We have%
\begin{equation}
\sigma _{\left( \mathrm{i}\right) }=\left(
\begin{array}{cccc}
a_{1} & 0 & c_{1} & 0 \\
0 & a_{1} & 0 & -c_{1} \\
c_{1} & 0 & a_{1} & 0 \\
0 & -c_{1} & 0 & a_{1}%
\end{array}%
\right) \qquad \sigma _{\left( \mathrm{ii}\right) }=\left(
\begin{array}{cccc}
a_{2} & 0 & c_{2} & 0 \\
0 & a_{2} & 0 & -c_{2} \\
c_{2} & 0 & a_{2} & 0 \\
0 & -c_{2} & 0 & a_{2}%
\end{array}%
\right) ,  \label{homogeneous}
\end{equation}%
\begin{equation}
\sigma _{\left( \mathrm{iii}\right) }=\left(
\begin{array}{cccc}
a_{1} & 0 & c_{3} & 0 \\
0 & a_{1} & 0 & c_{3} \\
c_{3} & 0 & a_{2} & 0 \\
0 & c_{3} & 0 & a_{2}%
\end{array}%
\right) \qquad \sigma _{\left( \mathrm{iv}\right) }=\left(
\begin{array}{cccc}
a_{1} & 0 & c_{4} & 0 \\
0 & a_{1} & 0 & -c_{4} \\
c_{4} & 0 & a_{2} & 0 \\
0 & -c_{4} & 0 & a_{2}%
\end{array}%
\right) .  \label{hybrid}
\end{equation}%
The matrices (\ref{homogeneous}-\ref{hybrid}) are of the form
\begin{equation}
\sigma _{\left( \mathrm{j}\right) }=\left(
\begin{array}{cc}
A_{_{\left( \mathrm{j}\right) }} & C_{_{\left( \mathrm{j}\right) }} \\
C_{_{\left( \mathrm{j}\right) }}^{T} & B_{_{\left( \mathrm{j}\right) }}%
\end{array}%
\right) ,  \label{Vj}
\end{equation}

\noindent with $\mathrm{j\in }$ subsystems$\left\{ \mathrm{(i)}\text{, }%
\mathrm{(ii)}\text{, }\mathrm{(iii)}\text{, }\mathrm{(iv)}\right\} $. The
covariance matrix $\sigma _{\left( \mathrm{j}\right) }$\ is real, symmetric
and positive and has block structure where $A_{\left( \mathrm{j}\right) }$, $%
B_{\left( \mathrm{j}\right) }$ and $C_{_{\left( \mathrm{j}\right) }}$ (and
its transpose) are $2\times 2$ Hermitian matrices. For each subsystem $%
\mathrm{j}$, $A_{_{\left( \mathrm{j}\right) }}$ and$\ B_{_{\left( \mathrm{j}%
\right) }}$ denote the symmetric covariance matrices for the individual
reduced one-mode states and the matrix $C_{_{\left( \mathrm{j}\right) }}$
comprise the cross-correlations between modes. The logarithmic negativity is
defined by \cite{VidalWerner30,Adesso et al32}
\begin{equation}
E_{N}^{\mathrm{(j)}}=\mathrm{max}\left\{ 0,-\ln (2\eta _{\left( \mathrm{j}%
\right) }^{-})\right\} ,  \label{negativity}
\end{equation}%
where $\eta _{\left( \mathrm{j}\right) }^{-}$ is the smallest simplistic
eigenvalue of the partial transpose of the $4\times 4$ covariance matrix $%
\sigma _{\left( \mathrm{j}\right) }$ \cite{Adesso et al32}. It writes
\begin{equation}
\eta _{\left( \mathrm{j}\right) }^{-}=\sqrt{\frac{\tilde{\Delta}_{_{\left(
\mathrm{j}\right) }}\text{ }-\text{ }\sqrt{\tilde{\Delta}_{_{\left( \mathrm{j%
}\right) }}^{2}-4\det \sigma _{_{\left( \mathrm{j}\right) }}}}{2}},
\label{Sympli HV}
\end{equation}%
where the symbol $\tilde{\Delta}_{_{\left( \mathrm{j}\right) }}$ stands for
the symplectic invariant for the covariance matrix (\ref{Vj}). It is given
by \cite{Adesso et al32}
\begin{equation*}
\tilde{\Delta}_{_{\left( \mathrm{j}\right) }}=\det A_{_{\left( \mathrm{j}%
\right) }}+\det B_{_{\left( \mathrm{j}\right) }}-2\det C_{_{\left( \mathrm{j}%
\right) }}.
\end{equation*}%
\noindent Using the results (\ref{system M}) and (\ref{system O}), one gets
\begin{equation}
2\eta _{\left( \mathrm{i}\right) }^{-}=2\left( a_{1}-c_{1}\right) =\frac{%
1+2n_{\mathrm{th}}}{1+\alpha }\bigg(\frac{1}{1+\beta }+\alpha \bigg)+\frac{%
\beta e^{-2r}}{(1+\alpha )(1+\beta )},
\end{equation}%
\begin{equation}
2\eta _{\left( \mathrm{ii}\right) }^{-}=2\left( a_{2}-c_{2}\right) =\frac{%
1+2n_{\mathrm{th}}}{1+\alpha }\frac{\alpha \beta }{1+\beta }+\frac{e^{-2r}}{%
1+\alpha }\left( \frac{\alpha }{1+\beta }+1\right) .
\end{equation}

\noindent For the subsystems $\mathrm{(iii)}$ and $\mathrm{(iv)}$. The
expressions of $2\eta _{\left( \mathrm{iii}\right) }^{-}$ and $2\eta
_{\left( \mathrm{iv}\right) }^{-}$ are too cumbersome and will not be
reported here. Clearly, the entanglement occurs when $E_{N}^{\mathrm{(j)}}>0$
or equivalently $\eta _{\left( \mathrm{j}\right) }^{-}<1/2$ with $\mathrm{%
j\in }$ subsystem$\left\{ \mathrm{(i)}\text{,}\mathrm{(ii)}\text{,}\mathrm{%
(iii)}\text{,}\mathrm{(iv)}\right\} $. The simplistic eigenvalues $\eta
_{\left( \mathrm{j}\right) }^{-}$ are a function of the squeezing parameter $%
r$, the opto-mechanical cooperativity $\beta $, the mean thermal photons
number $n_{\mathrm{th}}$ or equivalently the thermal bath temperature $T$
and the damping ratio $\alpha =\frac{\gamma }{\kappa }.$

\noindent As we shall hereafter focus on the difference between the
logarithmic negativity and the Gaussian quantum discord as quantifiers of
the quantum correlations, an appropriate choice of the parameters
characterizing the system is needed. In other hand, this must corresponds to
situations that can be implemented experimentally. In this respect, we
consider some parameters reported in \cite{GroeblacherVanner35}. The two
cavities are characterized by the length $L=25~\mathrm{\ mm}$, the laser
wave length is $\lambda =1064~\mathrm{nm}$, the frequency $\omega _{c}=2\pi
\times 5.26\times 10^{14}~\mathrm{Hz}$ and pumped by a coherent laser source
with power $P=11~\mathrm{mW}$. The movable mirrors having the mass $\mu =145~%
\mathrm{ng}$ and oscillate at the frequency $\omega _{\mu }=2\pi \times
947\times 10^{3}~\mathrm{Hz}$ with the mechanical damping rate $\gamma =2\pi
\times 140~\mathrm{Hz}$.

\begin{figure}[htb]
\centering
\begin{minipage}[htb]{2.8in}
\centering
\includegraphics[width=2.8in]{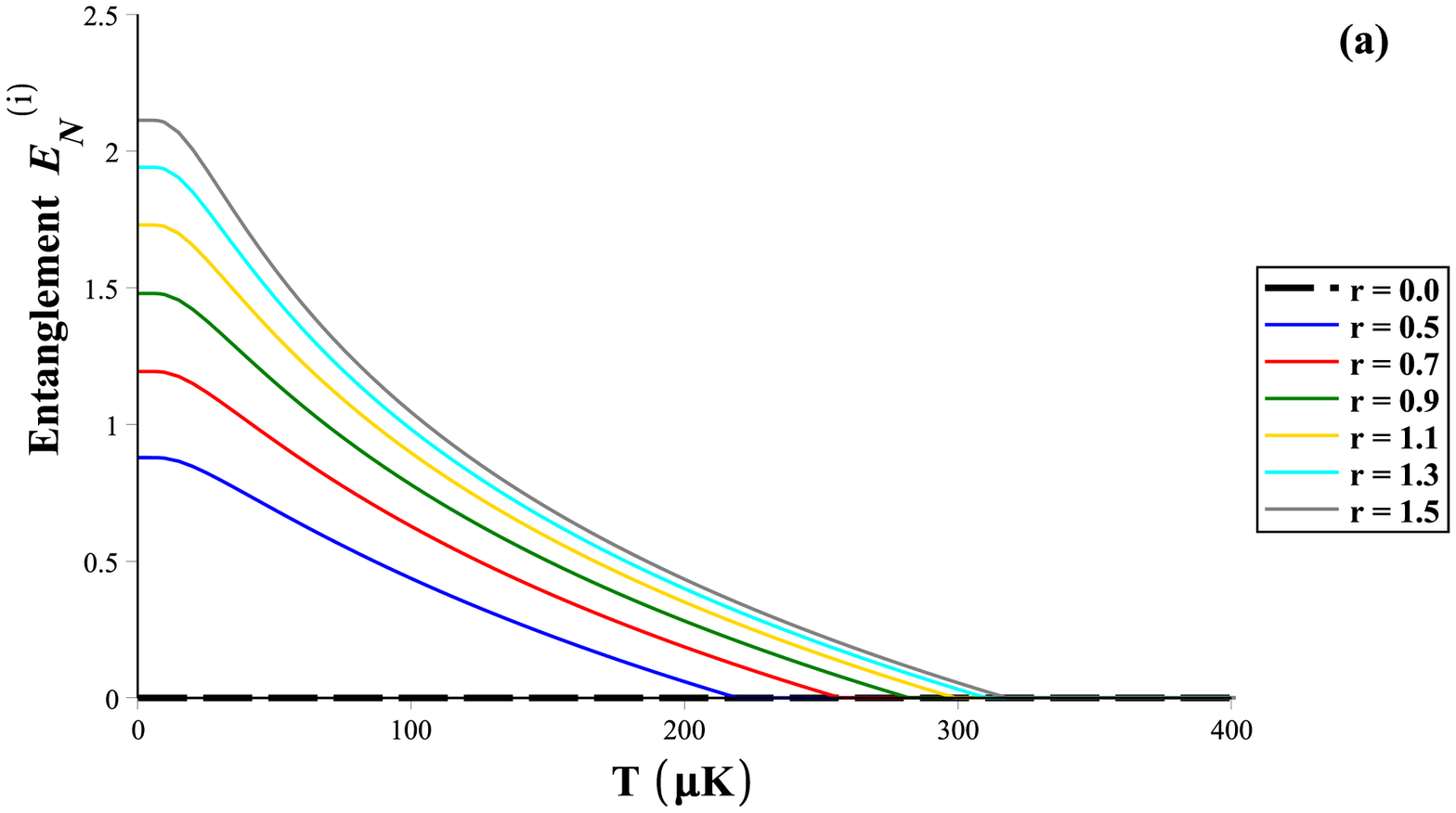}
\end{minipage}
\begin{minipage}[htb]{2.8in}
\centering
 \includegraphics[width=2.8in]{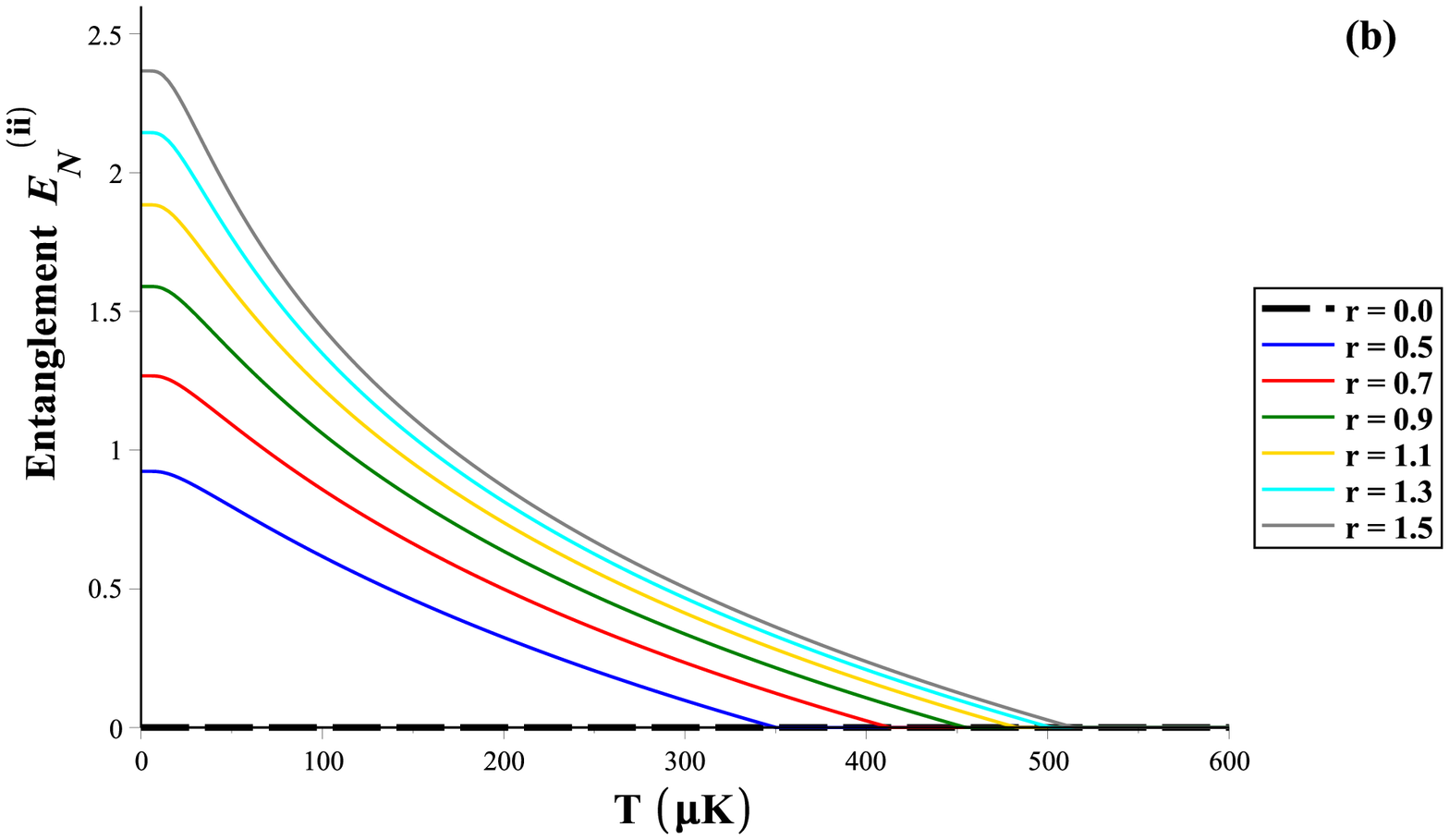}
\end{minipage}
\caption{ The logarithmic negativity $E_{N}$ versus the thermal bath
temperature $T$ for various values of the squeezing parameter $r$. (a): the
logarithmic negativity $E_{N}^{\mathrm{(i)}}$ of the subsystem $\mathrm{(i)}$
formed by two identical mechanical modes, (b): the logarithmic negativity $%
E_{N}^{\mathrm{(ii)}}$ of the subsystem $\mathrm{(ii)}$ composed by two
identical optical modes. For both cases (a) and (b), the opto-mechanical
cooperativity $\protect\beta$ is taken equal to $34$, the damping ratio $%
\protect\alpha =\frac{\protect\gamma }{\protect\kappa }$ is fixed to $0.05$
(or equivalently $\protect\kappa=2\protect\pi\times 2800~\mathrm{Hz}$).}
\end{figure}

Fig.2 shows that the logarithmic negativity $E_{N}^{\mathrm{(i)}}$ (resp. $%
E_{N}^{\mathrm{(ii)}}$) for the subsystems $\mathrm{(i)}$ (resp. $\mathrm{%
(ii)}$) decreases when the thermal bath temperature $T$ increases. In
particular, it is clearly seen that the logarithmic negativity $E_{N}^{%
\mathrm{(i)}}$ vanishes more quickly than $E_{N}^{\mathrm{(ii)}}$ under the
temperature effects. We notice also that, for a fixed value of the thermal
bath temperature, the quantities $E_{N}^{\mathrm{(i)}}$ and $E_{N}^{\mathrm{%
(ii)}}$ increase as the squeezing parameter increases. We remark that in the
absence of the squeezed light ($r=0$), the two mechanical modes of the
subsystem $\mathrm{(i)}$ and the two optical modes of the subsystem $\mathrm{%
(ii)}$ remain separable (see the black dashed lines in Figs.2(a) and 2(b)).
This reflects the relationship between the entanglement and the squeezed
light explains the quantum correlations transfer from squeezed light to
subsystems $\mathrm{(i)}$ and $\mathrm{(ii)}$ in agreement with the results
obtained in \cite{Sete et al}. From Fig.2 we also see that when the
squeezing parameter $r$ increases, the critical value of the thermal bath
temperature denoted $T_{0}$, from which the subsystems $\mathrm{(i)}$ and $%
\mathrm{(ii)}$ become separable decreases. The temperature $T_{0}$ is given
by
\begin{equation}
\frac{1}{T_{0}^{\left( \mathrm{i}\right) }}=\frac{k_{B}}{\hbar \omega _{\mu }%
}\ln \bigg(\frac{2(1+\alpha +\alpha \beta )}{\beta \left( 1-e^{-2r}\right) }%
+1\bigg),  \label{T-mechanical threshold}
\end{equation}%
for the subsystem $\mathrm{(i)}$ formed by the two mechanical modes. For the
case of the optical modes (subsystem $\mathrm{(ii)}$), it writes
\begin{equation}
\frac{1}{T_{0}^{\left( \mathrm{ii}\right) }}=\frac{k_{B}}{\hbar \omega _{\mu
}}\ln \bigg(\frac{2\alpha \beta }{(1+\alpha +\beta )(1-e^{-2r})}+1\bigg).
\label{T-optical threshold}
\end{equation}%
For the optical modes (the subsystem $\mathrm{(ii)}$), the logarithmic
negativity is more resilient against the temperature effects in comparison
with the mechanical modes (the subsystem $\mathrm{(i)}$). In fact, the
logarithmic negativity is zero beyond $T=3\times 10^{-4}~\mathrm{K}$ (for
the mechanical modes) and $T=5\times 10^{-4}~\mathrm{K}$ (for the optical
modes) regardless the value of the squeezing parameter. Such a phenomenon is
regularly known as entanglement sudden death (ESD) \cite{Daoud33,AlKasimi34}%
.
\begin{figure}[tbh]
\centering
\begin{minipage}[htb]{2.8in}
\centering
\includegraphics[width=2.8in]{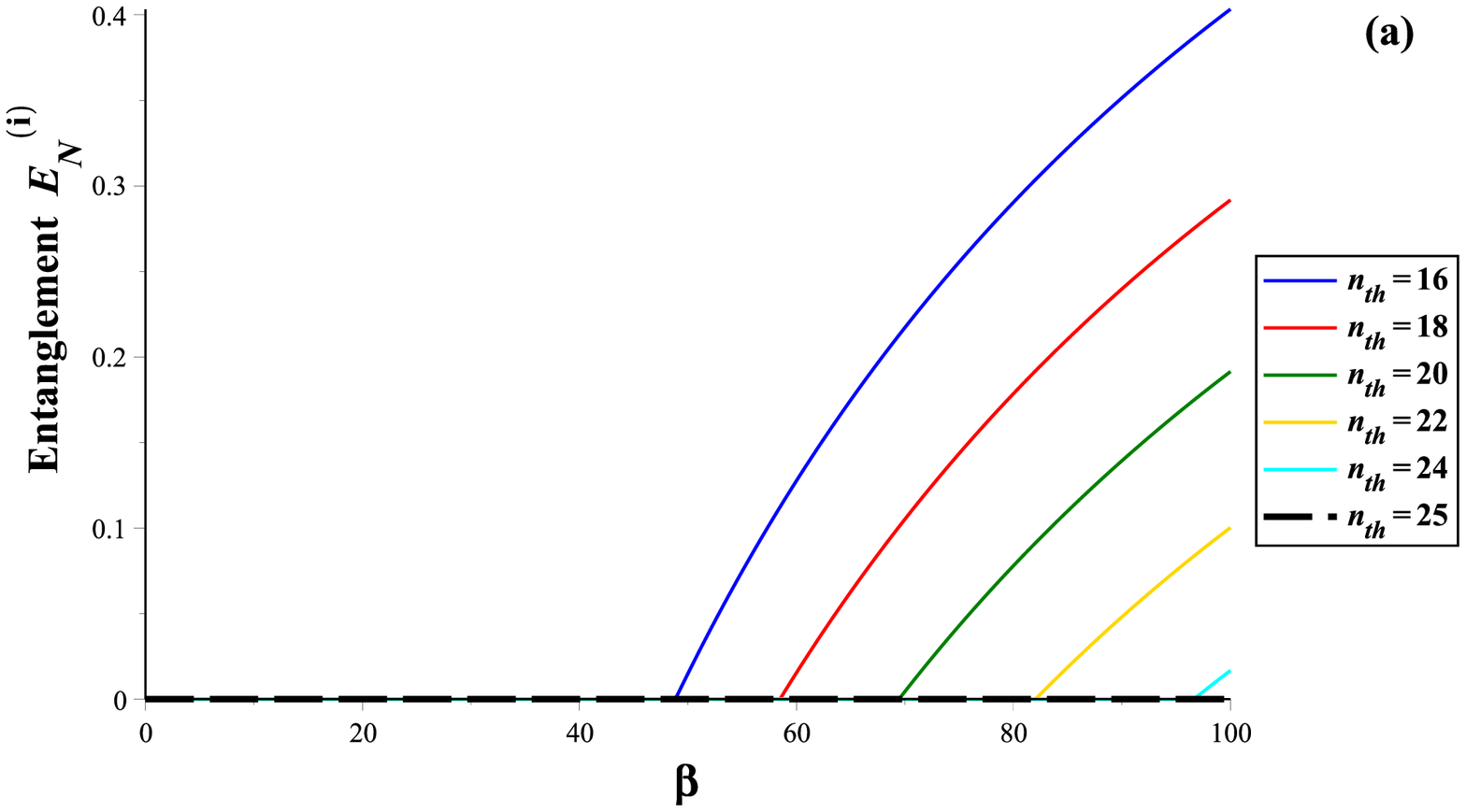}
\end{minipage}
\begin{minipage}[htb]{2.8in}
\centering
 \includegraphics[width=2.8in]{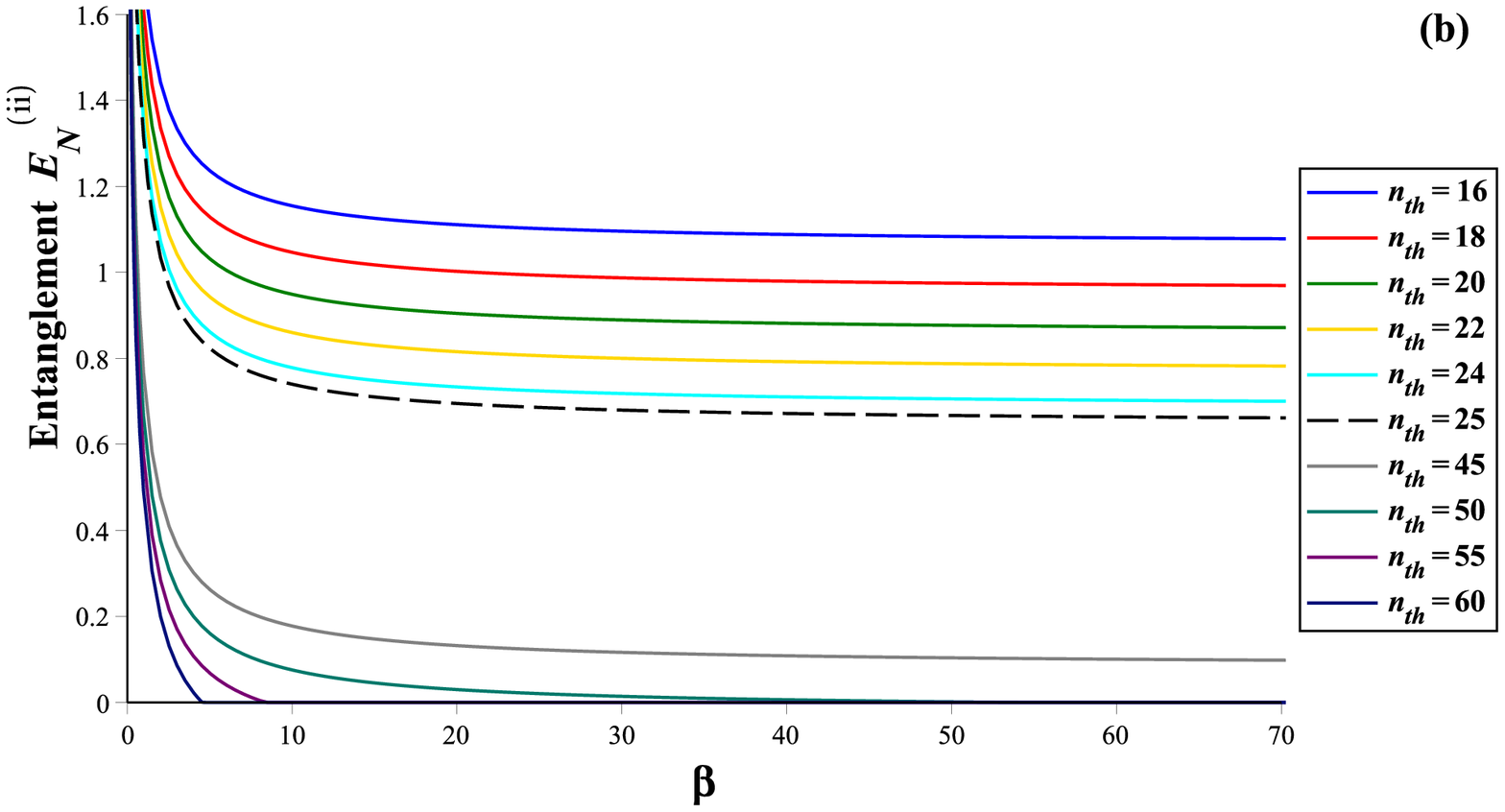}
\end{minipage}
\caption{Plots of the logarithmic negativity $E_{N}$ versus the
opto-mechanical cooperativity $\protect\beta $ for various values of the
mean thermal photons number $n_{\mathrm{th}}$. (a): the logarithmic
negativity $E_{N}^{\mathrm{(i)}}$ of the subsystem $\mathrm{(i)}$ formed by
two identical mechanical modes, (b): the logarithmic negativity $E_{N}^{%
\mathrm{(ii)}}$ of the subsystem $\mathrm{(ii)}$ composed by two identical
optical modes. In the two cases (a) and (b), we used $\protect\alpha =\frac{%
\protect\gamma }{\protect\kappa }=0.01$ (or equivalently $\protect\kappa =2%
\protect\pi \times 14\times 10^{3}~\mathrm{Hz}$). The squeezing parameter r
is fixed as $r=2$.}
\end{figure}
In Fig.3 we give the evolution of the logarithmic negativity $E_{N}^{\left(
\mathrm{i}\right) }$ and $E_{N}^{\left( \mathrm{ii}\right) }$ versus the
opto-mechanical cooperativity $\beta $ \ for various values of the mean
thermal photons number $n_{\mathrm{th}}$. The mechanical modes exhibit
vanishing the logarithmic negativity for $n_{\mathrm{th}}>25$. For $n_{%
\mathrm{th}}<25$, when $\beta $ increases, the mechanical modes are
entangled. The threshold value of the cooperativity $\beta _{0}^{\left(
\mathrm{i}\right) }$ beyond which the mechanical modes cease to be
separable, is given by
\begin{equation}
\beta _{0}^{\left( \mathrm{i}\right) }=\frac{2n_{\mathrm{th}}\left( 1+\alpha
\right) }{1-2\alpha n_{\mathrm{th}}-e^{-2r}}.  \label{C-mechanical threshold}
\end{equation}%
The optical modes remain entangled for $n_{\mathrm{th}}<25$ regardless the
value taken by the cooperativity. However, for higher values of $n_{\mathrm{%
th}}$ ($n_{\mathrm{th}}>25$ comparing with the case of the mechanical
modes), they start to be separable above the critical value $\beta
_{0}^{\left( \mathrm{ii}\right) }$ given by
\begin{equation}
\beta _{0}^{\left( \mathrm{ii}\right) }=\frac{\left( e^{-2r}-1\right) \left(
1+\alpha \right) }{1-2\alpha n_{\mathrm{th}}-e^{-2r}}.
\label{C-optical threshold}
\end{equation}%
Clearly, by increasing the mean thermal photons number $n_{\mathrm{th}}$,
the mechanical modes require a large value of $\beta $ to switch from
separable states to entangled states (see Fig.2(a)). This behavior can be
explained by the decoherence phenomenon. Indeed, $n_{\mathrm{th}}$ increases
when the thermal bath temperature $T$ increases and therefore the
environment effect on the system becomes more aggressive. Unlike the
mechanical modes (see Fig.2(a)), $E_{N}^{\left( \mathrm{ii}\right) }$
diminishes when $\beta $ increases (see Fig.2(b)). We now consider the case
of the hybrid opto-mechanical systems, formed by an optical cavity mode and
a mechanical mode. We start analyzing the hybrid subsystem $\mathrm{(iii)}$
which is composed by two interacting modes (an optical cavity mode and its
corresponding mechanical mode). In this case, we have $\det C_{\left(
\mathrm{iii}\right) }$= $\left( c_{3}\right) ^{2}>0$ and subsequently,
according to \cite{Simon,AdessoFabre} ($\det C<0$ is a necessary condition
for a two-mode Gaussian state to be entangled), the states of the two modes
forming the hybrid subsystem $\mathrm{(iii)}$ are always separable.
\begin{figure}[tbh]
\centerline{\includegraphics[width=7cm]{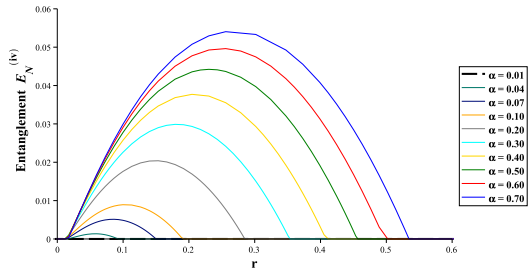}}
\caption{ The logarithmic negativity $E_{N}^{\mathrm{(iv)}}$ of the hybrid
subsystem $\mathrm{(iv)}$ formed by two uncoupled modes (an optical cavity
mode and non corresponding mechanical mode) versus the squeezing parameter $%
\mathrm{r}$ for different values of the damping ratio $\protect\alpha $. The
mean thermal photons number $n_{\mathrm{th}}$ and the opto-mechanical
cooperativity $\protect\beta $ are taken equal to $0.01$ and $1$
respectively.}
\end{figure}
Different entanglement behavior is obtained for the hybrid subsystem $%
\mathrm{(iv).}$ The results are reported in Fig.4. The logarithmic
negativity $E_{N}^{\left( \mathrm{iv}\right) }$ is depicted as a function of
the squeezing parameter r for various values of the damping ratio $\alpha $.
In the absence and also for small values of the squeezing parameter $r$,
Fig.4 reveals that no entanglement between the two modes forming the
subsystem $\mathrm{(iv)}$. This indicates that the squeezed light is a
necessary element to achieve entanglement. Such a result traduces the
transfer of the quantum correlations from the squeezed light to the
subsystem $\mathrm{(iv),}$ which agrees with the results obtained in
Figs.2(a) and 2(b). Fig.4 shows a resonant behavior of the entanglement $%
E_{N}^{\left( \mathrm{iv}\right) }$ in term of the squeezing parameter $r$.
The maximum value of $E_{N}^{\left( \mathrm{iv}\right) }$ increases with
increasing values of the damping ratio $\alpha .$ It must be noticed that
for a fixed value of $\alpha ,$ the entanglement $E_{N}^{\left( \mathrm{iv}%
\right) }$ is enhanced when $r$ increases before the resonance. This is no
longer valid after passing the resonant value of $E_{N}^{\left( \mathrm{iv}%
\right) }$. Indeed, for higher values of $r$, the entanglement goes to zero.
This is mainly due to thermal noise (affected each cavity) enhanced by
strong squeezing light as obtained in \cite{Mauro21}.

\section{Gaussian quantum discord}

In this section, we shall investigate the usefulness of the Gaussian quantum
discord \cite{GeordaParis9,AdessoDatta10} in comparison with the logarithmic
negativity discussed in the previous section. In particular, we shall focus
on the situations, discussed in section III, where the logarithmic
negativity is zero. Hence, using the Gaussian quantum discord defined in
\cite{GeordaParis9,AdessoDatta10}, we evaluate the quantum correlations
present in different subsystems $\mathrm{(i)}$, $\mathrm{(ii)}$, $\mathrm{%
(iii)}$ and $\mathrm{(iv)}$ at the separable states. For the bipartite
subsystem $\mathrm{j}$ ($\mathrm{j\in }$ subsystems$\left\{ \mathrm{(i)}%
\text{, }\mathrm{(ii)}\text{, }\mathrm{(iii)}\text{, }\mathrm{(iv)}\right\} $%
) described by the covariance matrix $\sigma _{\left( \mathrm{j}\right) }$
(Eq. (\ref{Vj})), the Gaussian quantum discord is given by \cite%
{GeordaParis9,AdessoDatta10}%
\begin{equation}
D^{\mathrm{(j)}}=f\Big(\sqrt{\det B_{\mathrm{(j)}}}\Big)-f\Big(\nu _{+}^{%
\mathrm{(j)}}\Big)-f\Big(\nu _{-}^{\mathrm{(j)}}\Big)+f\Big(\varepsilon ^{%
\mathrm{(j)}}\Big),  \label{GQD}
\end{equation}

\noindent where the function $f$ is defined by $f(x)=(x+\frac{1}{2})\log
_{2}(x+\frac{1}{2})-(x-\frac{1}{2})\log _{2}(x-\frac{1}{2}).$ The symplectic
eigenvalues $\nu _{+}^{\mathrm{(j)}}$ and $\nu _{-}^{\mathrm{(j)}}$ are
defined by \cite{GeordaParis9,AdessoDatta10}

\begin{equation}
\nu _{\pm }^{\mathrm{(j)}}=\sqrt{\frac{\Delta _{\mathrm{(j)}}\pm \sqrt{%
\Delta _{\mathrm{(j)}}^{2}-4\det \sigma _{\mathrm{(j)}}}}{2},}
\label{The symplectic eigenvalues}
\end{equation}%
with $\Delta _{\mathrm{(j)}}=\det A_{\mathrm{(j)}}+\det B_{\mathrm{(j)}%
}+2\det C_{\mathrm{(j)}}$. For the bipartite subsystems $\mathrm{(i)}$, $%
\mathrm{(ii)}$ and $\mathrm{(iv)}$ described respectively by the covariance
matrices $\sigma _{\mathrm{(i)}},\sigma _{\mathrm{(ii)}}$ and $\sigma _{%
\mathrm{(iv)}}$, $\varepsilon ^{\mathrm{(j)}}$ takes the following form \cite%
{GeordaParis9}

\begin{equation}
\varepsilon ^{\mathrm{(j)}}=\frac{\sqrt{\det A_{\mathrm{(j)}}}+2\sqrt{\det
A_{_{\mathrm{(j)}}}\det B_{_{\mathrm{(j)}}}}+2\det C_{_{\mathrm{(j)}}}}{1+2%
\sqrt{\det B_{_{\mathrm{(j)}}}}},  \label{Epsilon1,2,4}
\end{equation}

\noindent with $\mathrm{j\in }$ subsystem $\left\{ \mathrm{(i)}\text{,}%
\mathrm{(ii)}\text{,}\mathrm{(iv)}\right\} $. For the subsystem $\mathrm{%
(iii)}$ defined by the matrix $\sigma _{\left( \mathrm{iii}\right) }$, we
have $C_{\mathrm{(iii)}}=$diag($c_{3},+c_{3}$). Then $\varepsilon ^{\mathrm{%
(j)}}$ is given by the formula \cite{AdessoDatta10,Olivares}

\begin{equation}
\varepsilon ^{\mathrm{(j)}}=\frac{2\left\vert \det C_{\mathrm{(j)}%
}\right\vert +\sqrt{4\left( \det C_{\mathrm{(j)}}\right) ^{2}+(4\det B_{%
\mathrm{(j)}}-1)(4\det \sigma _{\mathrm{(j)}}-\det A_{\mathrm{(j)}})}}{%
\left( 4\det B_{\mathrm{(j)}}-1\right) },  \label{Epsilon3}
\end{equation}%
with $\mathrm{j\equiv (iii)}$.
\begin{figure}[tbh]
\centering
\begin{minipage}[htb]{2.8in}
\centering
\includegraphics[width=2.8in]{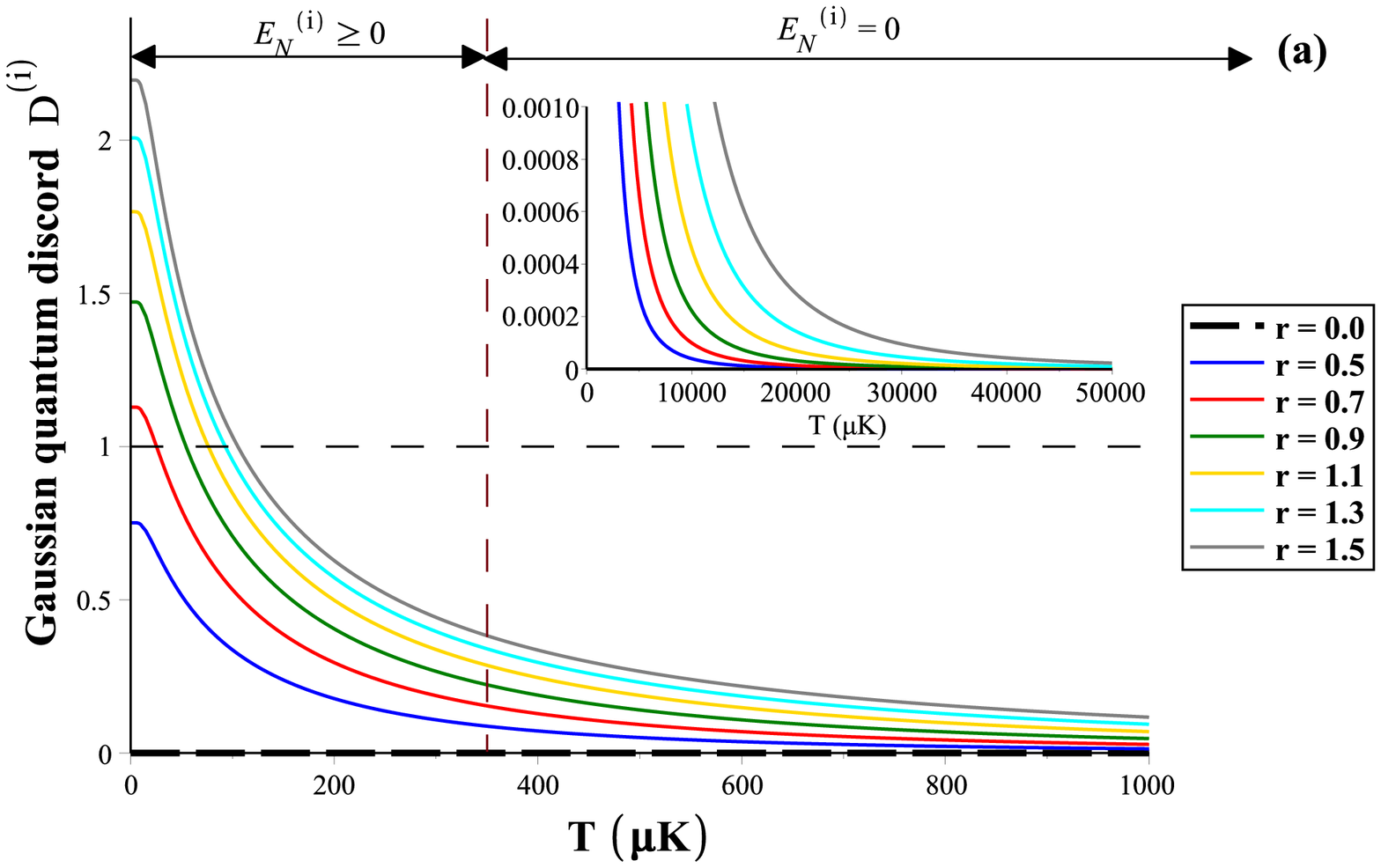}
\end{minipage}
\begin{minipage}[htb]{2.8in}
\centering
 \includegraphics[width=2.8in]{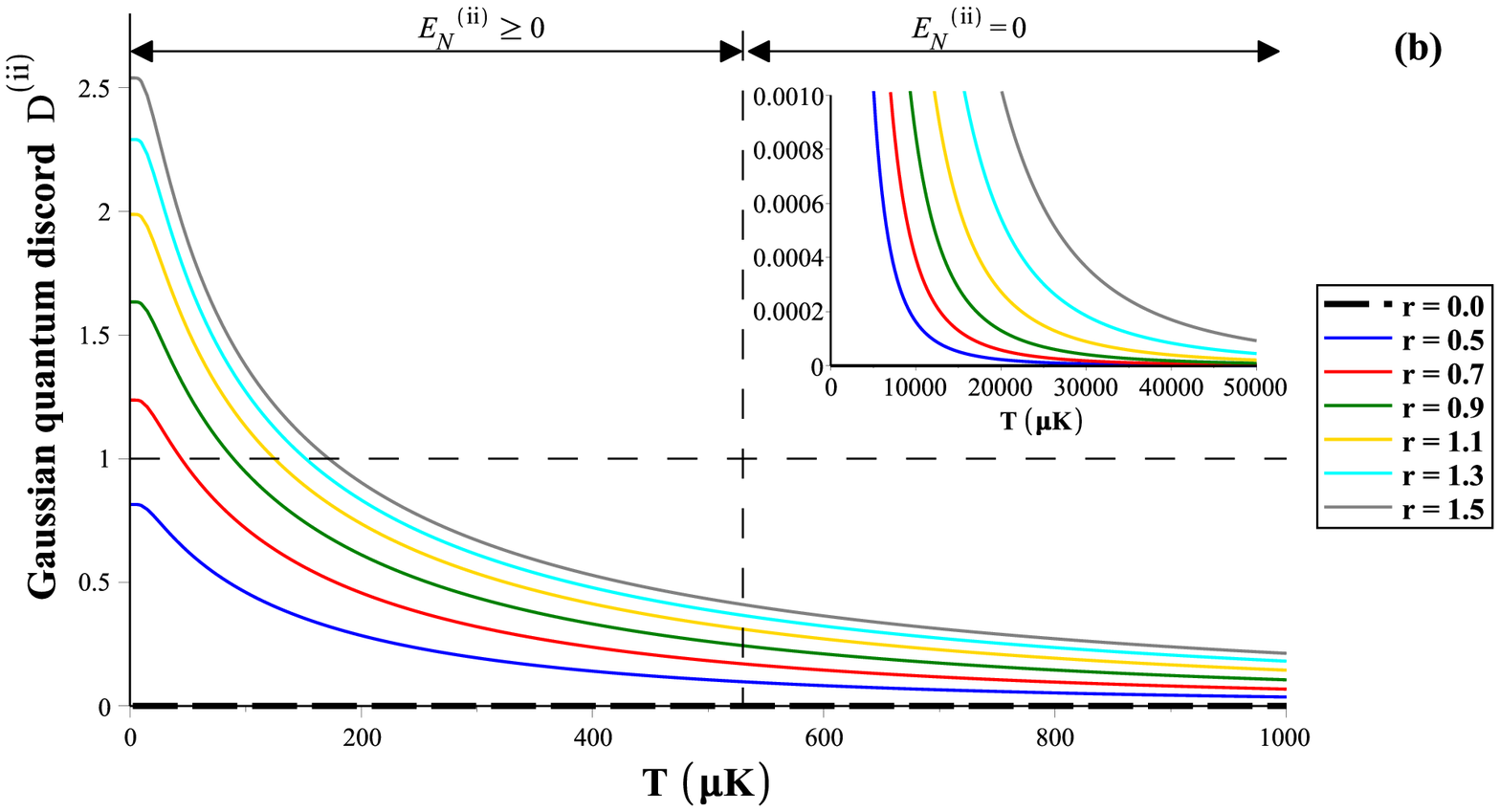}
\end{minipage}
\caption{ Plots of the Gaussian quantum discord $D$ against the thermal bath
temperature $T$ for various values of the squeezing parameter $r$. (a): the
Gaussian quantum discord $D^{\mathrm{(i)}}$ of the subsystem $\mathrm{(i)}$,
(b): the Gaussian quantum discord $D^{\mathrm{(ii)}}$ of the subsystem $%
\mathrm{(ii)}$. For both cases (a) and (b), the parameters $\protect\alpha =%
\frac{\protect\gamma }{\protect\kappa }$ and $\protect\beta $ are fixed as
the same as in Fig.2. The vertical dashed lines show the boundary between
the separable states ($E_{N}=0$) and entangled states ($E_{N}\neq 0$).
Figs.(5a) and (5b) show that the Gaussian quantum discord presents non-zero
values in the same regions where the subsystems $\mathrm{(i)}$ and $\mathrm{%
\ (ii)}$ are separable (see Figs.2(a) and 2(b)), which is an indicator of
quantumness of correlations in the considered subsystems (existence of
non-classical correlations even at the separable states).}
\end{figure}
Having the necessary ingredients to deal with the Gaussian quantum discord
for the various subsystems of the opto-mechanical system under
consideration, we investigate firstly the quantum correlations measured by $%
D^{\left( \mathrm{i}\right) }$ and $D^{\left( \mathrm{ii}\right) }$ present
respectively in the homogeneous subsystems $\mathrm{(i)}$ and $\mathrm{(ii).}
$ In Fig.5 we give the variations of $D^{\left( \mathrm{i}\right) }$ and $%
D^{\left( \mathrm{ii}\right) }$ as function of the thermal bath temperature $%
T$ for different values of the squeezing parameter $r$. The damping ratio $%
\alpha $ and the opto-mechanical cooperativity $\beta $ take the same values
as in Fig.2. Clearly, the quantum discord $D^{\left( \mathrm{i}\right) }$
(for the mechanical modes) and $D^{\left( \mathrm{ii}\right) }$ (for the
optical modes) decrease when the thermal bath temperature increases. Using
the results reported in Figs.(2a) and (2b) we notice that the logarithmic
negativity vanishes beyond $T\approx 3.5\times 10^{-4}K$ for the mechanical
modes $(E_{N}^{\left( \mathrm{i}\right) }=0)$ and beyond $T\approx
4.75\times 10^{-4}K$ for the optical modes $(E_{N}^{\left( \mathrm{ii}%
\right) }=0)$. However, for the already mentioned ranges of thermal bath
temperature, the Gaussian quantum discord $D^{\left( \mathrm{i}\right) }$
and $D^{\left( \mathrm{ii}\right) }$ are non zero . This indicates that the
Gaussian quantum discord measure seems more robust and resilient versus the
effect of the environment (decoherence) and constitutes a good tool to
decide about the existence of non-classical correlations (quantumness) in
opto-mechanical systems. This result corroborate the fact that quantum
correlations exist in the subsystems $\mathrm{(i)}$ and $\mathrm{(ii)}$ even
at the separable states. We note also that, when $E_{N}^{\left( \mathrm{j}%
\right) }=0$ we have $D^{\left( \mathrm{j}\right) }<1$ with $\mathrm{j\in }$
subsystem $\left\{ \mathrm{(i)}\text{,}\mathrm{(ii)}\right\} $, which is
agrees with the analysis reported in \cite{GeordaParis9,AdessoDatta10}.
Another important aspect, we investigate in this paper, concerns the
behavior of the Gaussian quantum discord of the mechanical and optical modes
(the subsystem $\mathrm{(i)}$ and $\mathrm{(ii)}$) in terms of the
opto-mechanical cooperativity $\beta $. This is reported in Fig.6. The
Gaussian quantum discord is plotted as a function of the opto-mechanical
cooperativity $\beta $ for various values of the mean thermal photons number
$n_{\mathrm{th}}$. The damping ratio $\alpha $ and the squeezing parameter $r
$ are fixed as in Fig.3 in order to compare the logarithmic negativity with
the Gaussian quantum discord as quantifiers of quantum correlations in the
subsystems $\mathrm{(i)}$ and $\mathrm{(ii)}$.
\begin{figure}[tbh]
\centering
\begin{minipage}[htb]{2.8in}
\centering
\includegraphics[width=2.8in]{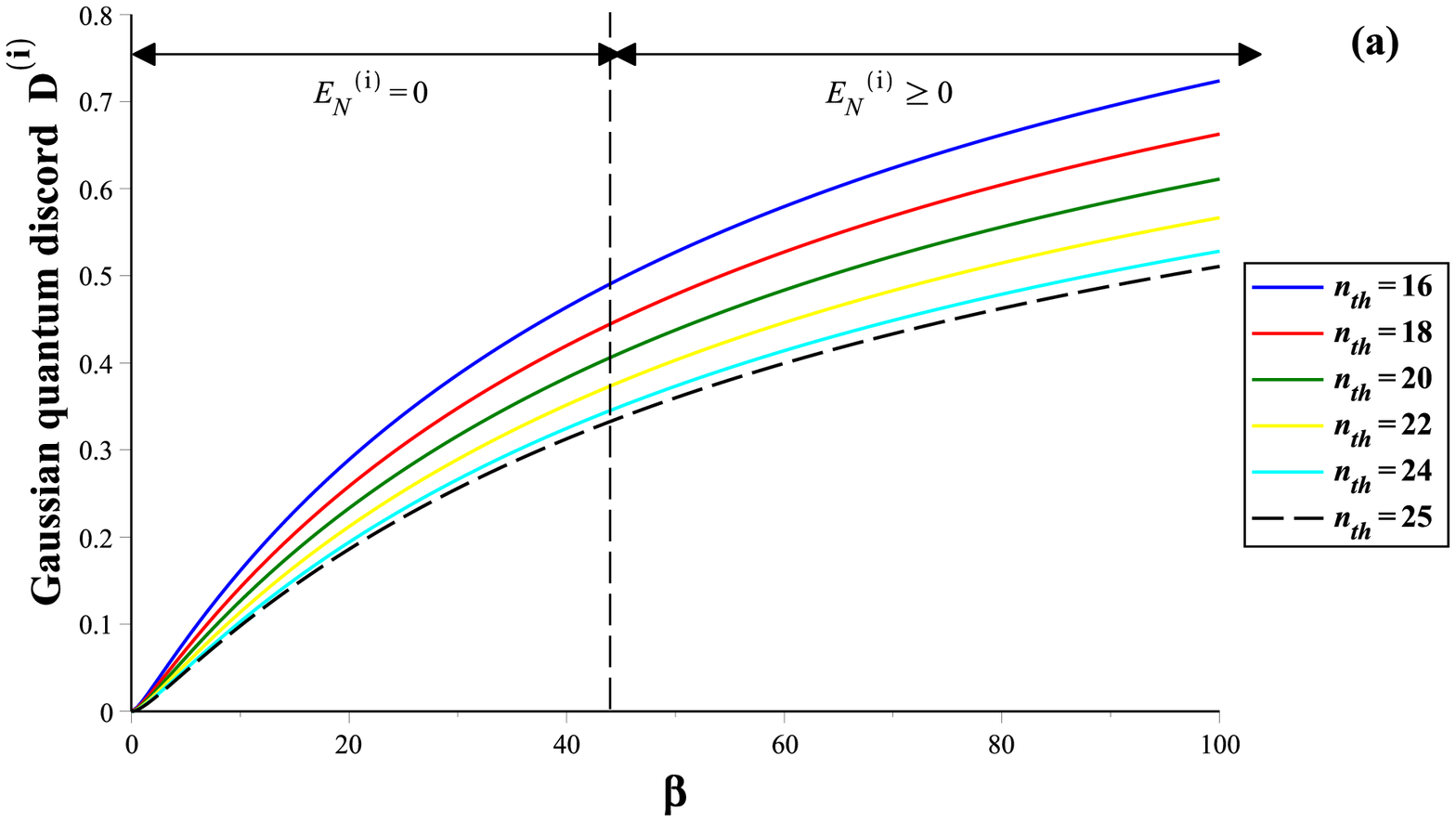}
\end{minipage}
\begin{minipage}[htb]{2.8in}
\centering
 \includegraphics[width=2.8in]{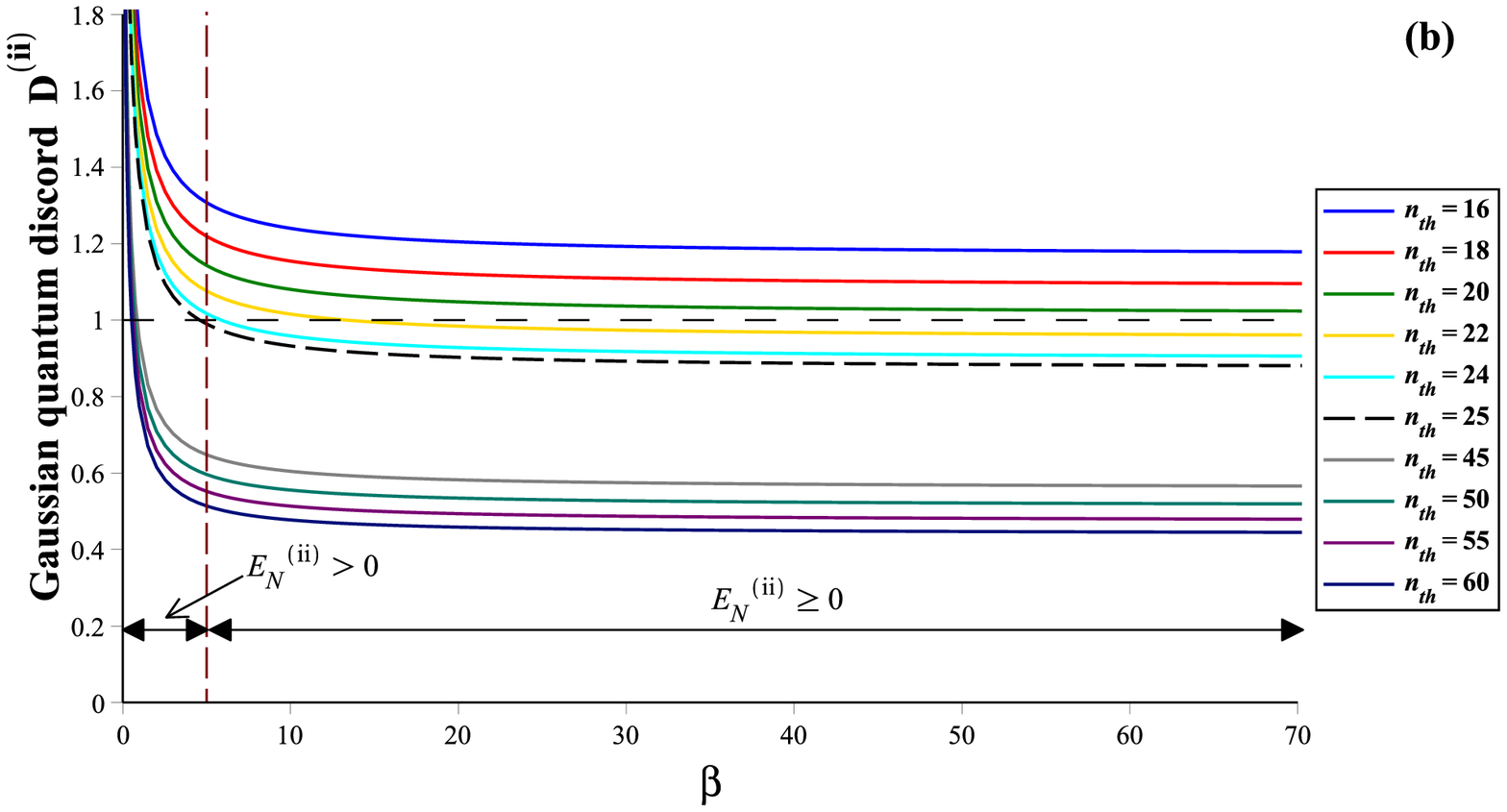}
\end{minipage}
\caption{ The Gaussian quantum discord $D$ versus the opto-mechanical
cooperativity $\protect\beta $ for various values of the mean thermal
photons number $n_{\mathrm{th}}$. Panel (a) shows the Gaussian quantum
discord $D^{\mathrm{(i)}}$ of the subsystem $\mathrm{(i)}$, panel (b) shows
the Gaussian quantum discord $D^{\mathrm{(ii)}}$ of the subsystem $\mathrm{%
(ii)}$. The parameters $\protect\alpha $ and $r$ are fixed as the same as in
Fig.3. The vertical dashed lines show the boundary between, separable and
entangled states. It is easy to remark that: for the subsystem $\mathrm{(i)}$%
, when $\protect\beta \in \left[ 0,50\right] $, $E_{N}^{\mathrm{(i)}}=0$ and
$D^{\mathrm{(i)}}\neq 0$ (see Figs.(3a) and (6a)), concerning the subsystem $%
\mathrm{(ii)}$ and focusing on the case where $n_{\mathrm{th}}=60$, we can
see that, for $\protect\beta >5$, $E_{N}^{\mathrm{(ii)}}=0$ and $D^{\mathrm{%
(ii)}}\neq 0$ (see Figs.3(b) and 6(b)). Therefore, such situations, make
sure the existence of quantumness of correlations between the two modes
formed the subsystems $\mathrm{(i)}$ and $\mathrm{(ii)}$.}
\end{figure}
The Gaussian quantum discord $D^{\left( \mathrm{i}\right) }$ increases with
increasing values of the cooperativity $\beta $\ (see Fig.6(a)) but this
increasing becomes slow for higher mean thermal photons number $n_{\mathrm{th%
}}$. In the other hand, the quantum discord $D^{\left( \mathrm{ii}\right) }$
decreases as the cooperativity increases and becomes almost constant for
higher values of $\beta $. It must be also noticed that for the non
separable optical modes, the diminution of the quantum discord is more
pronounced for higher thermal photons number $n_{\mathrm{th}}$. Comparing
Figs.6(a) and 6(b) we deduce that there is a tradeoff of the intricacy
between the optical and mechanical modes. Indeed, for small values of the
cooperativity $\beta $, the mechanical modes are separable while the optical
modes are not and by increasing the cooperativity, the mechanical modes
become non separable and the optical modes are separable.
\begin{figure}[tbh]
\centering
\begin{minipage}[htb]{2.9in}
\centering
\includegraphics[width=2.9in]{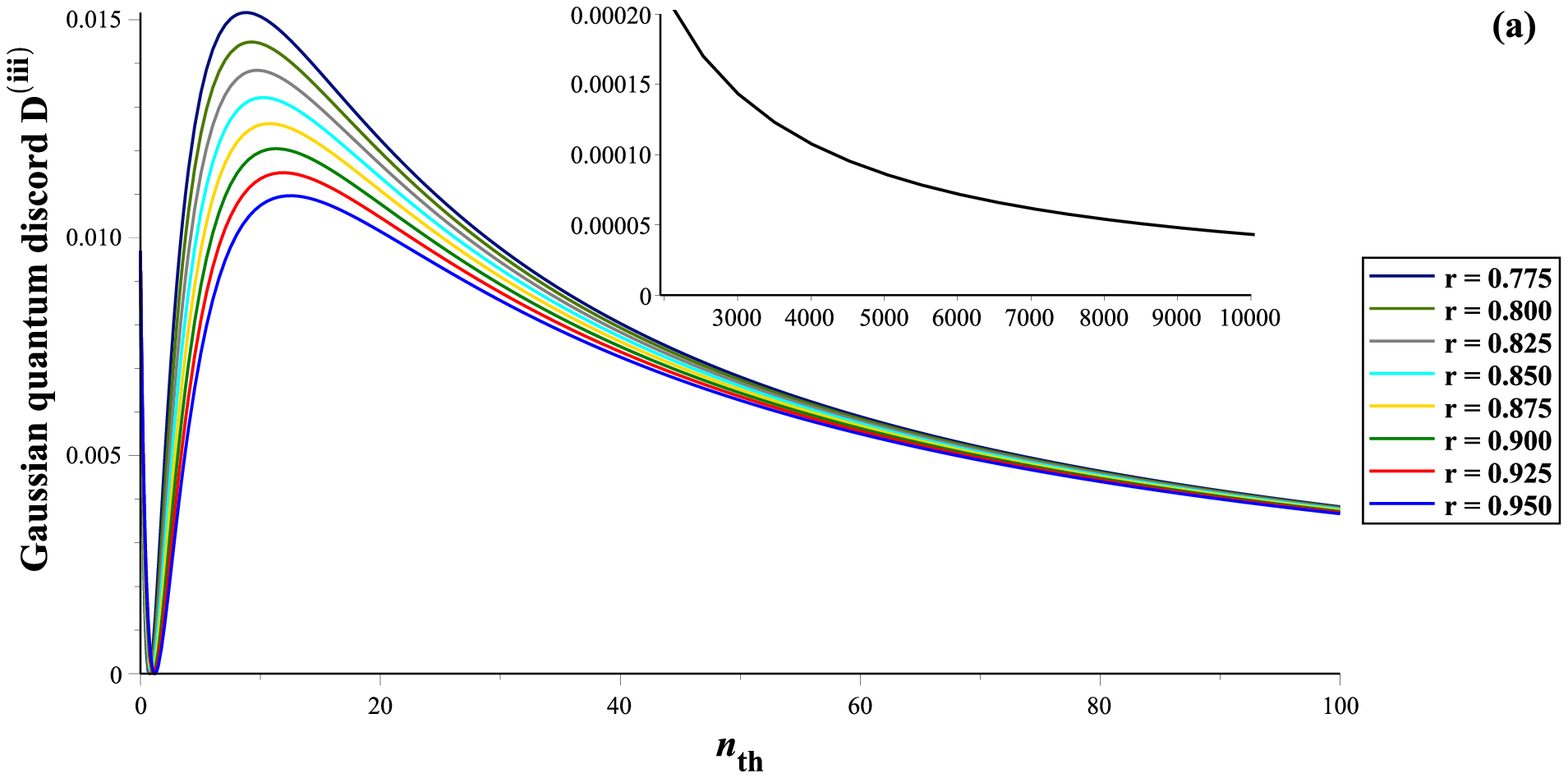}
\end{minipage}
\begin{minipage}[htb]{2.9in}
\centering
 \includegraphics[width=2.9in]{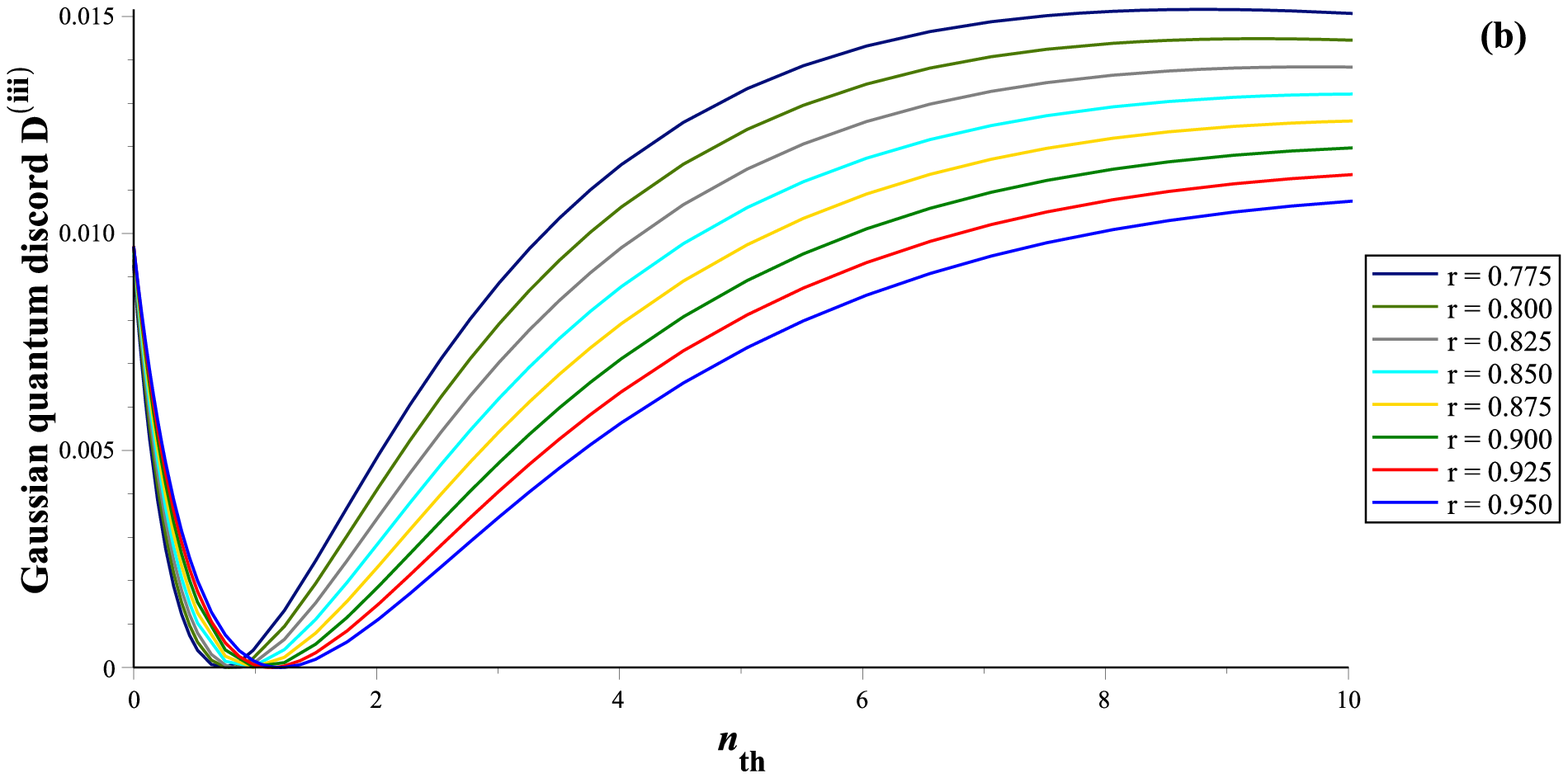}
\end{minipage}
\caption{The Gaussian quantum discord $D^{\mathrm{(iii)}}$ of the hybrid
subsystem $\mathrm{(iii)}$ formed by two interacting modes (an optical
cavity mode and its corresponding mechanical mode) versus the mean thermal
photons number $n_{\mathrm{th}}$ ((a) behavior for high values of $n_{\mathrm{th}}
$, (b) behavior for small values of $n_{\mathrm{th}}$) for different values of
the squeezing parameter $r$. The damping ratio $\protect\alpha $ and the
opto-mechanical cooperativity $\protect\beta $ \ are taken equal to $0.5$
and $10$ respectively.}
\end{figure}
Now, we consider the Gaussian quantum discord in the hybrid subsystem $%
\mathrm{(iii).}$ The robustness of the Gaussian quantum discord $D^{\left(
\mathrm{iii}\right) }$ with respect to the mean thermal photons number $n_{%
\mathrm{th}}$ (or equivalently the thermal bath temperature $T$) for various
values of the squeezing parameter $r$ is shown in Fig.7. This figure shows
that $D^{\left( \mathrm{iii}\right) }$ has two distinct behaviors according
to $n_{\mathrm{th}}$. Indeed, for small values of $n_{\mathrm{th}}$ ($0<n_{%
\mathrm{th}}<1$) and for a given value of $r$, $D^{\left( \mathrm{iii}%
\right) }$ decreases quickly from a non zero initial value, reaching a
minimum around $n_{\mathrm{th}}\approx 1$ (see Fig.7(b)), whereas for $n_{%
\mathrm{th}}>1$, $D^{\left( \mathrm{iii}\right) }$ has a resonant behavior
(the maximums decrease when $r$ increase and attained around $n_{\mathrm{th}%
}\approx 10$ (see Fig.7(a)). Finally, it is clear that $D^{\left( \mathrm{iii}%
\right) }$ remains non zero for high values of $n_{\mathrm{th}}$ ($n_{%
\mathrm{th}}>10^{4}$) and keeps a value almost constant independently of $r$
(see Fig.7(a)). We recall that the subsystem $\mathrm{(iii)}$ is always
separable and the Gaussian quantum discord $D^{\left( \mathrm{iii}\right) }$
is less than 1. This is in agreement with the general properties of Gaussian
quantum discord \cite{GeordaParis9,AdessoDatta10}. Therefore the quantum
correlations detected in this situation are a witness of quantumness.
\begin{figure}[tbh]
\centerline{\includegraphics[width=7cm]{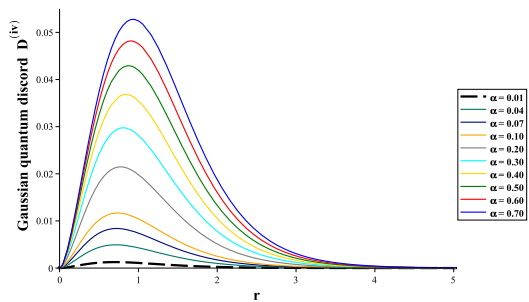}}
\caption{ The Gaussian quantum discord $D^{\mathrm{(iv)}}$ of the hybrid
subsystem $\mathrm{(iv)}$ formed by an optical cavity mode and non
corresponding mechanical mode versus the squeezing parameter $\mathrm{r}$
for different values of the damping ratio $\protect\alpha $. The mean
thermal photons number $n_{\mathrm{th}}$ and the opto-mechanical
cooperativity $\protect\beta $ are taken equal to $0.01$ and $1$
respectively. }
\end{figure}
\noindent The behavior of the Gaussian quantum discord $D^{\left( \mathrm{iv}%
\right) }$ of the hybrid subsystem $\mathrm{(iv)}$ is plotted as a function
of the squeezing parameter $r$ in Fig.$8$, various values of the damping
ratio $\alpha $ are considered. Fig.$8$ shows that the Gaussian quantum
discord is non zero for $0<r\leq 0.02$ and $r>0.55$ where the logarithmic
negativity is zero (see Fig.$4$), in this case, we have also $D^{\left(
\mathrm{iv}\right) }<1$. Finally, using the standard homodyne detection
method, it is possible to determine numerically the global covariance matrix
( Eq. (\ref{Global CM})) by the measure of the correlations between the
output fields, which provides an experimental method to quantify stationary
entanglement and Gaussian quantum discord by means of Eqs. (\ref{GQD}) and (%
\ref{negativity}). More technical details are presented in Refs.\cite%
{VitaliVedral17, AdessoFabre},

\section{Concluding Remarks}

To summarize, we have investigated the quantum correlations in a quantum
opto-mechanical system describing the interaction between light and
mechanical systems in a Markovian environment without the adiabatic
approximation. We considered an opto-mechanical system consisting by two
identical Fabry-Perot cavities. We gave the quantum Langevin equations (see
Eqs. (\ref{cdote}) and (\ref{bdote})) from which we derived the dynamics of
the optical as well as the mechanical degrees of freedom. A crucial feature
is that all the quadratures of optical and mechanical modes are expanded to
the first-order around the steady states (see Eqs. (\ref{operat-fluct}) and (%
\ref{c and b average}) ). In this picture, the quantum Langevin equation
gives a coupled system of differential equations involving noise operators
(see Eqs. (\ref{deltacdote}) and (\ref{deltabdote})). Our analysis is not
very different from other proposals discussed recently in the literature.
Differences become relevant when we incorporate in the model the quantum
correlations in various bipartite subsystems (four subsystems). Indeed,
given an arbitrary steady state, the fluctuations about it are fully
characterized by its $8\times 8$ covariance matrix of all pairwise
correlations among the quadratures. To compute pairwise correlations, we
used the $4\times 4$ sub-matrices given by Eqs. ((\ref{homogeneous}),(\ref%
{hybrid})) which are extracted from the global covariance matrix $\sigma$
(see Eq. (\ref{Global CM})). They correspond to the four subsystems $\mathrm{%
(i)}$, $\mathrm{(ii)}$, $\mathrm{(iii)}$ and $\mathrm{(iv)}$. The covariance
sub-matrix $\sigma _{\left( \mathrm{i}\right) }$ (resp. $\sigma _{\left(
\mathrm{ii}\right) }$) (see Eq. (\ref{homogeneous})) associated to the
homogeneous subsystem $\mathrm{(i)}$ (resp. $\mathrm{(ii)}$ ) describes the
correlations between the mechanical (resp. optical) modes. On the other
hand, the covariance sub-matrices $\sigma _{\left( \mathrm{iii}\right) }$
and $\sigma _{\left( \mathrm{iv}\right) }$ (see Eq. (\ref{hybrid}))
associated with the hybrid subsystem $\mathrm{(iii)}$ and $\mathrm{(iv))}$
contain the information about the quantum correlations between the
mechanical and optical modes in the opto-mechanical system under
consideration. This global description allows us to access to the
non-classical correlations existing between each pair of the quadrature
components. In evaluating the pairwise correlations, we deliberately
considered the logarithmic negativity which characterizes the degree of
entanglement and the Gaussian quantum discord which quantifies the
non-classical correlations not captured by entanglement. A particular focus
was devoted to states with vanishing logarithmic negativity (separable
states) for which the Gaussian discord is non zero. We have depicted the
opto-mechanical entanglement evolution under the thermal bath temperature,
the opto-mechanical cooperativity, the squeezing parameter of the light and
the mean thermal photons number. The results, reported in Figs.2(a) and
2(b), show that the entanglement between the optical modes (see Fig.2(b))
are more robust against the temperature effects than the mechanical modes
(see Fig.2(a)). Furthermore, from Figs.2(a) and 2(b), it is clear that the
squeezed light enhances the entanglement between the optical modes
(subsystem $\mathrm{(ii)}$) and the mechanical modes (subsystem $\mathrm{(i)}
$) especially for lower thermal bath temperatures. In the subsystem $\mathrm{%
(iv)}$, the logarithmic negativity is quadratic in term of the squeezing
parameter (i.e.$~E_{N}^{\mathrm{(iv)}}~\sim ~r^{2}$). This indicates that to
attain the maximal value of the correlations between the optical and
mechanical modes in the subsystem $\mathrm{(iv)}$, one has to choose a
special value of the squeezing parameter. The Gaussian quantum discord in
the subsystem $\mathrm{(i)}$ and $\mathrm{(ii)}$ follows rigourously the
same behavior in terms of the temperature. It is important to notice that
for the subsystems $\mathrm{(i)}$, $\mathrm{(ii)}$ and $\mathrm{(iv)}$,
which are formed by two spatially separable modes, it is indispensable to
use the squeezed light to create entanglement and Gaussian quantum discord.
This indicates the quantum correlations transfer from the squeezed light to
the two considered modes. In the subsystem $\mathrm{(iii)}$, the mechanical
mode and the optical mode are always separable (the logarithmic negativity
is zero) but the corresponding pairwise quantum correlation is non zero when
measured by Gaussian quantum discord. More interesting, for moderates values
of mean thermal photons number $\mathrm{n}_{th}$, the Gaussian quantum
discord tends to an asymptotic constant value. This constitutes a very
interesting and at the same time surprising result. Indeed, in the subsystem
$\mathrm{(iii)}$, it seems that low thermal effect enhances the quantum
correlations. Our results confirm the robustness of the Gaussian quantum
discord, in comparison with the entanglement, for the four partitions $%
\mathrm{(i)}$, $\mathrm{(ii)}$, $\mathrm{(iii)}$ and $\mathrm{\ (iv)}$
comprised in the opto-mechanical system investigated in this paper.

\end{document}